\documentclass[12pt,thmsa]{article}
\usepackage{amsfonts}

\begin{document}

\author{Karl-Georg Schlesinger \\
e-mail: kggschles@gmail.com}
\title{Entropy, heat, and G{\"o}del incompleteness}
\date{22.02.2014}
\maketitle

\begin{abstract}
Irreversible phenomena -- such as the production of entropy and heat -- arise from fundamental reversible dynamics because the forward dynamics is too complex, in the sense that it becomes impossible to provide the necessary information to keep track of the dynamics. On a heuristic level, this is well captured by coarse graining. We suggest that on a fundamental level the impossibility to provide the necessary information might be related to the incompleteness results of G{\"o}del. This would hold interesting implications for both, mathematics and physics.

\end{abstract}

\bigskip

Ever since Boltzmann tried to reduce the laws of thermodynamics in a statistical sense to the laws of mechanics, it has been a puzzle how the notions of entropy and heat of thermodynamics -- which are fundamentally tied to \textit{irreversible} processes in physics -- could emrge from the time reversible laws of classical mechanics. The explanations invoked either presuppose irreversibility by explicit or hidden assumptions (e.g. claiming that probabilistic arguments are only to be applied to future events while past events are \textit{facts}, see e.g. \cite{Wei}) or rely on a form of coarse graining. Definitely, the latter is an efficient calculational tool but as a \textit{fundamental} explanation it would reduce something as the production of heat to an arbitrary choice of coarse description. It does not help to call on the fundamental limits of measurement in quantum mechanics at this point: Quantum mechanics does not limit precision in configuration space but only in phase space. In other words, quantum mechanics only limits the precision to be achieved for coordinate \textit{and} monentum at the same time, for neither of the two separately. At the same time, the dynamics of quantum mechanics is described by a first order differential equation (in contrast to the second order of classical mechanics) which just means that it is completely sufficient to have only one variable (i.e. the coordinate) measured precisely, in order to have sufficient initial conditions to determine the dynamics. And this dynamics is, once again, time reversible, i.e. it does not offer any obvious explanation for irreversible physics. We can also not refer to the collapse of the wavefunction in measurement since the production of heat is not dependent on the stage at which we interfer with our observation (this stage -- or cut -- in the description can be nearly arbitrarily delayed, as shown by von Neumann, see \cite{Neu}).\\
The reversibility of the fundamental mechanical laws remains true even if we pass to quantum field theory (neither does the inclusion of general relativity for a description of gravity change the situation since it is itself positioned on the level thermodynamics, i.e. it does not provide a microscopic explanation for e.g. black hole entropy). Sometimes it is argued that finally the Planck length does provide a fundamental coarse graining in confuguration space. But a Planck scale cut-off relates to ultra high energies and it is not easily seen how this would match the energy scales found e.g. in the production of heat by friction. More fundamentally, AdS/CFT and black holes provide a lot of theoretical evidence for the concept of holography, i.e. quantum gravity would be a unitary theory and, therefore, time reversible, again (Strictly speaking, it does only imply that there is always a unitary description for a suitable choice of reference frame, e.g. that of the outside observer at infinity for black holes). \\
It is true, though, that -- on a heuristic basis -- coarse graining \textit{does} cover the effect. What coarse graining does is giving precision -- as a calulational tool -- to our naive perception that processes in physics which are irreversible have a complicated forward dynamics. At some point, the dynamics becomes so complicated, then, that the coarse grained description can no longer deliver sufficient information. So, it is the \textit{complexity} of the dynamics which should offer the clue to an explanation of irreversibility. \\

\bigskip

Let us take a look at a much simpler situation, in the form of the concept of a universal computing machine (Turing machine). As is well known, in this case, we only have a nonvanishing contribution to entropy (and a production of heat) if information is erased. The computational steps are all reversible. Now, erasing informations means that the information is lost \textit{forever}. The concept of a Turing machine is an axiomatic concept, i.e. it does not provide any explanation how this erasing is to be achieved in the physical world. But it shows that the erasing of information has to be \textit{fundamental} in order to lead to a production of entropy and heat. We can not think of this erasing as a lack of information due to finite computational power. If we can recover the information by a more powerful computer (even if it has to be more powerful on a cosmological magnitude), there is no production of entropy or heat. The reason is that the Turing machine is a \textit{universal} computing machine, and only if the information is erased in the sense of this \textit{universal} machine, do we get a production of entropy and heat. \\
The concept of coarse graining is just the concept of a computer of finite computational power in disguise. So, the results on entropy and Turing machines give very strong -- and mathenatically rigorous -- doubt about using coarse graining in a \textit{fundamental} explanation for irreversibility. \\
But the case of universal Turing machines offers another clue. G{\"o}del's incompleteness results imply that there are problems which are \textit{fundamentally} beyond the reach of universal Turing machines (and, therefore, beyond mathematical acessability since mathematical axiom systems are nothing but programs -- or program languages -- running on a universal Turing machine). The results of \cite{Moo} show that this results in a lack of predictability of the dynamics of all sufficiently complex systems of differential equations. The results of \cite{GA} give evidence that this holds, especially, for the Einstein equations of general relativity. AdS/CFT suggests that that it should then also apply to certain quantum field theories. \\
If the dynamics of a system becomes so complex that G{\"o}del incompleteness prohibits a complete description of its dynamics, the necessary information -- to determine the dynamics -- is \textit{fundamentally} lost on a \textit{universal} Turing machine. This should -- from the results on universal Turing machines, mentioned above -- imply a production of entropy and heat. So, the results of \cite{Moo} offer the possibility for a new Ansatz which could lead to a fundamental understanding of irreversibility and the production of entropy and heat from G{\"o}del incompleteness for dynamical systems of sufficient complexity. \\

\bigskip

We should remark that this Ansatz could -- as does coarse graining if used heuristically -- only work for the usual case of -- on a cosmological scale -- isolated physical systems (like the case of friction). It can \textit{not} explain why, on a cosmological scale, we make the experience that much more systems have a more complicated forward dynamics as vice versa. The mentioned Ansatz might only explain \textit{how} entropy is produced \textit{if} the forward dynamics is sufficiently complex. The cosmological observation is rooted in the fact that the universe started from highly ordered -- and \textit{extremely} unlikely -- initial conditions. An explanation for this can only be searched for in a deeper understanding of the beginning of the universe. \\

\bigskip

A few remarks are important at this point:

\begin{itemize}

\item If one considers \textit{true} dynamical systems, described by differential equations -- instead of the axiomatic concept of a universal Turing machine -- the non-predictability of the dynamics depends on the choice of axioms system $A$ which one uses to formalize the basis of mathematics (typically, $A$ would be the Zermelo-Fraenkel axiom system of set theory). Of course, a phenomenon, as the production of heat, can not depend on the choice of $A$. Formulated differently, the choice of $A$ corresponds to a choice of program (or program language), running on a universal Turing machine. But if we could recover the dynamics by upgrading to better software (a different choice of axiom system $A$), this can not correspond to a production of entropy and heat (since the information is not truely erased from the universal Turing machine, in this case). So, we would need a slightly stronger form of G{\"o}del incompleteness which would make the dynamics non-predictable for \textit{any} choice of axiom system $A$. \\
The results of \cite{Nab} suggest that something anlong these lines might appear in general relativity. The results of \cite{Nab} show that for any \textit{fixed} choice of $A$, one can always find a covariance problem (equivalence problem of two manifolds) which can not be decided (since one can find a manifold which can homotopically not be decided from the sphere). Since manifolds of infinite genus are not excluded from the path integral in gravity, it looks like one should be able to find a relevant covariance problem in general relativity which would not be decidable for \textit{any} choice of axiom system.\\
Besides this, general relativity possesses a kind of universality, concerning complexity in dynamical systems: Any halting problem for universal Turing machines can be translated into a covariance problem in general relativity (see \cite{Nab}). So, it might be possible that \textit{any} theory, involving a production of entropy and heat by sufficiently complex dynamics (e.g. effective limits of quantum field theories, as relevant for friction), has a \textit{dual} gravitational description (along the lines of e.g. the Navier-Stokes/gravity duality). If this would hold, one could reduce a proof, for the production of entropy and heat as a consequence of G{\"o}del incompleteness, to the case of general relativity.

\item G{\"o}del incompleteness has a very clear description in terms of complexity. We can attach a degree of complexity to any choice of axiom system $A$. If the dynamics of a system of differential equations becomes non-predictable, we can understand this as the dynamics of the system becoming too complex, relative to the complexity of $A$ (see \cite{CC}, \cite{Cha}). The entropy should then be a quantitative measure of how much the complexity of the dynamics exceeds that of $A$, i.e. it should relate to the complexity of the dynamics \textit{relative} to the complexity of $A$. Here, two steps are important: First, to translate to a complexity measure relevant to physical systems, there can be a nontrivial scaling factor involved if one starts from the complexity introduced in \cite{CC} and \cite{Cha}. One has to fix this, e.g. by studying a suitable example. Secondly, it is obviuous that the -- in this way suitably scaled -- complexity of axiom sytems has to be bounded if one lets it run over all axiom sytems $A$. Otherwise it would be impossible to extract a finite entropy from dynamics which should be too complex for any axiom system $A$ (in the way just sketched in the previous item). \\
If the above idea -- that it might be possible to reduce the general question to the case of general relativity -- would hold true, the needed determination of the scaling factor could be tried e.g. for a black hole solution (like the Kerr solution). The entropy would then have to be derived from the complexity of the dynamics of the underlying microscopic degrees of freedom. Taking e.g. black hole entropy and considering these microscopic degrees of freedom in the context of string theory, the task would essentially reduce to questions of computational complexity/G{\"o}del incompleteness for the Calabi-Yau landscape. This is a setting where complexity considerations are known to play a role (see \cite{DD}, \cite{Den}). \\
In this setting, it would be an interesting question if the mentioned bound for the -- suitably scaled -- complexity of $A$ relates to a special Calabi-Yau and -- if yes -- what its meaning could be. 

\item Considering black hole entropy from complexity of the microscopic degrees of freedom is somewhat different from studying G{\"o}del incompleteness directly for the classical Einstein equations (as in \cite{GA}). But black holes constitute the case of \textit{maximal} entropy in general relativity, we can always throw an entropic system into a black hole and increase its entropy further. In other words, \textit{all} entropic situations in general relativity can be absorbed by black holes which suggests that even in general relativity one might possibly be able to reduce the general case to the one of black holes. In this way, the case of \cite{GA} should be compatible with the microscopic explanation, e.g. from Calabi-Yau complexity (as one should expect from the standpoint of physics).

\end{itemize}

Finally, let us remark that rooting the production of entropy and heat in G{\"o}del incompleteness would -- if possible -- have interesting implications for both, mathematics and physics. On the side of physics, questions like the meaning of a possible complexity bounding Calabi-Yau (see above) would turn up. Also, the production of heat or its absence would turn into a powerful measure for computational unpredictability (or predictability) of dynamics. On the side of mathematics, it would imply that constructive approaches -- which try to circumvent the incompleteness results of G{\"o}del -- can never provide a mathematical basis for all of physics (since G{\"o}del incompleteness would be \textit{necessary} for true physical phenomena like the production of heat).

\end{document}